\documentclass[12pt]{article}
\usepackage{graphicx}
\usepackage{verbatim}

\input epsf
\def\beq{\begin{eqnarray}}
\def\eeq{\end{eqnarray}}

\def\ba{\begin{eqnarray}}
\def\ea{\end{eqnarray}}
\def\beq{\begin{eqnarray}}
\def\eeq{\end{eqnarray}}

\def\L*{{\cal L}_*}
\def\L{\mathcal{L}}
\def\({\left(}
\def\){\right)}

\def\lsim{\mathrel{\rlap{\lower3pt\hbox{\hskip0pt$\sim$}}
     \raise1pt\hbox{$<$}}}         
\def\gsim{\mathrel{\rlap{\lower4pt\hbox{\hskip1pt$\sim$}}
     \raise1pt\hbox{$>$}}}         

\def\lsim{\mathrel{\rlap{\lower3pt\hbox{\hskip0pt$\sim$}}
     \raise1pt\hbox{$<$}}}         
\def\gsim{\mathrel{\rlap{\lower4pt\hbox{\hskip1pt$\sim$}}
     \raise1pt\hbox{$>$}}}         

\usepackage{amsmath}
\usepackage{amsfonts}
\usepackage{verbatim}

\begin{document}

\begin{center}


{\Large\bf  Safety of  Minkowski  Vacuum}
 \vspace{0.2cm}

\end{center}

\begin{center}

{\bf Gia Dvali} 


\vspace{.2truecm}

\centerline{\em Arnold Sommerfeld Center for Theoretical Physics,
Fakult\"at f\"ur Physik} 
\centerline{\em Ludwig-Maximilians-Universit\"at M\"unchen,
Theresienstr.~37, 80333 M\"unchen, Germany}


{\em Max-Planck-Institut f\"ur Physik,
F\"ohringer Ring 6, 80805 M\"unchen, Germany}


{\em CERN,
Theory Division,
1211 Geneva 23, Switzerland}


{\em CCPP,
Department of Physics, New York University\\
4 Washington Place, New York, NY 10003, USA} \\

\end{center}



 

\centerline{\bf Abstract}

   We give a simple argument suggesting  that in a consistent quantum field theory tunneling from  Minkowski to  a lower energy  vacuum must be  impossible.   Theories that allow for such a tunneling also allow for localized states  of negative mass, and therefore, should be  inconsistent.


\vspace{0.5cm}

  The idea that we may be living in a false vacuum was a source of a lot of imagination about 
  the possible fate of our Universe. 
   Our vacuum is Minkowskian  with a great accuracy. 
   Thus, as a first step would be desirable to  better understand  the prospects  of tunneling 
   from the Minkowaski vacuum to a lower energy state. 
   
    A systematic  study of this issue was pioneered by Coleman and De Luccia \cite{cd}, who showed, that the 
   tunneling  goes through the formation of a vacuum bubble, and has non-zero probability 
   as long as the size of the critical bubble is finite. 
    They also pointed out that in certain cases,  although the lower vacuum does exist, tunneling 
    never takes place, because the effective size of the critical bubble is infinite. 
    This mechanism is referred to as Coleman-De Luccia suppression. 
    
     In this short note, we wish to point out that in any consistent quantum field theory tunneling 
   from a Poincare-invariant (Minkowski) vacuum   should never happen. This means that the lower energy vacuum  (if it exists) must be above the 
  Coleman-De Luccia bound. If this is not the case, the theory is inconsistent.  Thus, one way or the other,  a Poincare-invariant Minkowski  vacuum should  be stable by 
  consistency of the theory. 
  
     The argument for this stability is pretty simple.  In order to prove it, let us imagine the opposite, and get convinced that we will encounter  an  inconsistency.   
     So let  us imagine, that we are in a Minkowski vacuum from which tunneling to a lower energy vacuum is possible.  This means that a critical bubble of the true vacuum has a finite size.  
    Such a bubble of course has a zero  total energy.   However,  the finiteness of its size  means that (by continuity)  the same potential admits also other configurations that have negative total energy. 
  For any smooth scalar potential that admits a zero energy bubble of finite size, 
 the negative energy bubble  can always  be prepared by deforming the zero energy one.
  Of course, such a bubble will not be a static solution of the equations of motion, but 
  there is no need for this.  If such configuration exists, it will represent a localized object 
  of a  {\it negative } (ADM \cite{adm}) mass.  Any theory admitting such objects  is a disaster.   This is obvious already from the fact that 
   from large distance point of view  such a negative-mass bubble will look as a negative-mass 
   particle, and should share responsibility for all the trouble that such  particles cause. 
   
    To a reader that is still not convinced that having a negative mass objects is a  killer for a quantum field theory defined on a Poincare-invariant background, we  offer to perform the following thought experiment. 
    
    Consider  a decay of the Minkowski vacuum into a  negative-energy bubble (bubble$_-$)  plus some positive-energy  wave-packet  (bubble$_+$),    
 \begin{equation}
 {\rm  vacuum}  \, \rightarrow    \, {\rm bubble}_-  \,  +  \, {\rm bubble}_+  \, .
 \label{decay}
 \end{equation}     
  The role of the bubble$_+$  can be played  by  any positive-energy excitation in the theory, in particular by 
  a positive energy lump of the same scalar field that allows interpolation between Minkowski 
  and AdS vacua.  The problem is, that  in an interacting field theory that admits  localized  negative energy states  such a process is impossible to forbid.  Once not  forbidden,  it has an infinite rate, because due to Poincare-invariance the two lumps can be produced at an  arbitrary relative momenta. 
  
   It is the simplest to understand the physical meaning of this infinity  from the point of view of a 
   large-distance observer that is observing the  process at distances larger than  a characteristic 
 size of the lumps.  For such an observer process is seen as a pair-creation of negative and positive mass particles, with four-momenta $p_{\mu}$ and $q_{\mu}$. 
  By Poincare-invariance of the background, the  rate of pair-creation,  $\Gamma$,   can only depend  on invariant scalar products  $p^2$,  $q^2$ and $pq$. However  because of four-momentum conservation, which implies $p = -q$, all invariants are the same and equal to particle mass$^2$ ($ \equiv \, m^2$) .   Thus,  $\Gamma(p^2) \, = \, \Gamma(m^2)$ is 
    momentum-independent.   Consequently,  the total rate obtained by integrating over all final relative momenta  ($p\, - \, q \, = \, 2p$) is infinite.    
    
   The infinity appears because the pair has opposite masses and 
  thus the same velocities.  This infinity is physical.  
  Such  a divergent  summation over the relative momentum  can never take place in  theories where there are no negative energy states, since the final momenta are restricted by energy
  conservation/positivity.  
   
        Another way to see that  the divergent summation over the relative momentum is unavoidable, 
     is  to introduce a  Poincare-violating  preferred frame  parameterized  by an infinitesimal 
     vector $\epsilon_{\mu}$  and then take the limit $\epsilon \rightarrow 0$.    The role of 
    such a vector is to  set a "soft" reference frame.   The role of such a frame can be played e.g., by an uniform density of a spectator scalar particle of  a positive  mass  $\epsilon$, which can decay into the above pair of positive and negative energy states.   An uniform density of such particles then sets the preferred frame described by vector $\epsilon_{\mu}$.   Such uniform density can be taken to be arbitrarily small and  respectively  its effect on the decay can be made
as small as one wishes.  The vacuum decay is recovered in the zero density limit. 
  
      In the presence of such a frame, the  
    integration over relative momentum  is an  obvious necessity. 
     Of course,  for any non-zero $\epsilon_{\mu}$ the rate now can  also depend on the  the products $p_{\mu} \epsilon^{\mu}$,  but this dependence disappears  in the limit 
 $\epsilon = 0$.    
    
  
     We thus see, that the only  consistent Poincare-invariant  vacua are the exactly stable ones.  
 Lower  energy vacua either should not exist or be above the Coleman-De Luccia bound. 
 
  An immediate consistency check for our claim is to note that in supergravity the 
 Minkowski vacua,  which are known to allow only positive energy states \cite{sugra}, 
 are also stable under tunneling \cite{weinberg}.   
   But our arguments show that the stability of Minkowski vacuum is a matter of consistency 
 regardless of supersymmetry.   A consistent theory cannot 
be formulated on a metastable  Poincare-invariant vacuum.    Since our argument is based solely on the Poincare-invariance, the only possible loophole 
  would be to explicitly break it.  This possibility is beyond our interest (although see \cite{slava}).        
 
    Above reasoning  straightforwardly  applies and restricts other possible forms of scalar potentials that interpolate between the Minkowskli and AdS vacua with or without barriers. 
 
   Regarding our own vacuum,  although it is not exactly Minkowskian the same reasoning should apply, due to tiny difference.   Even if tunneling 
 can take place it will be rather soft.  So the vacuum stability is not  on the list of things that our civilization has to worry about.

  \vspace{02mm}
\centerline{\bf Acknowledgments}
It is a pleasure  to thank  Andrei Barvinsky,  Gaume Garriga, Cesar Gomez, Slava Mukhanov and Alex Vilenkin for discussions.
 This work was supported in part by Humboldt Foundation under Alexander von Humboldt Professorship,  by European Commission  under 
the ERC advanced grant 226371,   by TRR 33 \textquotedblleft The Dark
Universe\textquotedblright\   and  by the NSF grant PHY-0758032.


\begin{thebibliography}{99}
\bibitem{slava}   Such a Poincare-violation may be "self-imposed" 
due to finite life-time of the vacuum (private communications with S.~Mukhanov).   This would be consistent  with our results.
\end{thebibliography}
\end{document}